\documentclass[aps,prl,superscriptaddress,showpacs,amsmath,amssymb]{revtex4}
\usepackage{array}
\usepackage[dvips]{graphicx}
\usepackage{color}

\begin{document}

\title{Statistical patterns of visual search for hidden objects}

\author{Heitor F. Credidio}
\email{credidio@fisica.ufc.br}
\affiliation{Departamento de F\'{\i}sica, Universidade Federal
do Cear\'a, 60451-970 Fortaleza, Cear\'a, Brazil}

\author{Elis\^angela N. Teixeira}
\email{teixeiraelis@gmail.com }
\affiliation{Departamento de P\'os-Gradua\c{c}\~{a}o em Lingu\'istica,
Universidade Federal do Cear\'a, 60451-970 Fortaleza, Cear\'a, Brazil}

\author{Saulo D. S. Reis}
\email{saulo@fisica.ufc.br}
\affiliation{Departamento de F\'{\i}sica, Universidade Federal
do Cear\'a, 60451-970 Fortaleza, Cear\'a, Brazil}

\author{Andr\'e A. Moreira}
\email{auto@fisica.ufc.br}
\affiliation{Departamento de F\'{\i}sica, Universidade Federal
do Cear\'a, 60451-970 Fortaleza, Cear\'a, Brazil}

\author{Jos\'e S. Andrade Jr.}
\email{soares@fisica.ufc.br}
\affiliation{Departamento de F\'{\i}sica, Universidade Federal
do Cear\'a, 60451-970 Fortaleza, Cear\'a, Brazil}

\date{\today}

\begin{abstract}
  The movement of the eyes has been the subject of intensive research
  as a way to elucidate inner mechanisms of cognitive processes. A
  cognitive task that is rather frequent in our daily life is the
  visual search for hidden objects. Here we investigate through
  eye-tracking experiments the statistical properties associated with
  the search of target images embedded in a landscape of distractors.
  Specifically, our results show that the twofold process of eye
  movement, composed of sequences of fixations (small steps)
  intercalated by saccades (longer jumps), displays characteristic
  statistical signatures. While the saccadic jumps follow a log normal
  distribution of distances, which is typical of multiplicative
  processes, the lengths of the smaller steps in the fixation
  trajectories are consistent with a power-law distribution. Moreover,
  the present analysis reveals a clear transition between a
  directional serial search to an isotropic random movement as the
  difficulty level of the searching task is increased.
\end{abstract}

\maketitle

It is a common misconception to believe that memories are stored in
the brain the same way a movie is stored in a hard drive. Remembering,
just like seeing and listening, is in fact an act of construction much
more complex than usually thought, where vasts amounts of information
are processed and interpreted by the brain in order to create what we
call memories, and pretty much everything else we call reality
\cite{Neisser1967}. The field dedicated to the study of these types of
processes is called Cognitive Science, which took its current form in
the first half of the 20th century out of a mishmash of sciences,
including, among others, Psychology, Linguistics and Computer Science.
More precisely, the main challenge of the Cognitive Science is to
answer questions related to the way in which the brain processes
available information and how this shapes behaviour.
\cite{kintsch1984}.

As theoretical entities, cognitive processes cannot be directly
observed and measured \cite{Heyes2000}. Thus, in order to be able to
study them, we need to rely on observations about the behaviour of
individuals. A very often utilized approach is to follow the eye
movement during cognitive tasks. By the end of the XIX century, it was
still thought that the eyes smoothly scanned the line of text during
reading.  Louis \'Emile Javal~\cite{Javal2010}, in his unprecedented
study of 1879, observed that the eyes actually move in a succession of
steps, called fixations, followed by jerk-like movements, called
saccades, that are too fast to capture new visual information
\cite{Underwood1998a}. The method of eye-tracking as a fundamental
source of information about cognition was finally introduced through
the seminal work of Yarbus \cite{Yarbus1967}. This study provided
unambiguous demonstration for the fact that the movement of the eyes
is strongly correlated with the cognitive objectives of the
individual.

A cognitive process that benefits the most from the study of eye
movement is the visual search for hidden objects
\cite{Liversedge2000}, like when trying to find a person in a crowded
place, or a 2 inches nail inside a box of nails of various sizes.  An
early theory related to this process is due to Treismann and Gelade,
called Feature Integration Theory (FIT) \cite{Treisman1980}. This
theory deals with attention, a kind of mental focus that can be
directed to a desired region of the visual scene, therefore enhancing
the perceptual sensitivity in that region.  The FIT proposes that
visual search tasks are divided into two stages.  The first is a
detection stage, in which a small set of simple separable features
like color, size and orientation are identified in the elements inside
the optical array. This stage is a preattentive one, that is,
attention need not be directed at each element of the image in order
to perform detection, all feature registration takes place in parallel
across the whole visual scene.  In the second stage, called
integration, the features identified in the previous stage are
combined in order to conceive more complex characteristics.  This is
an attentive stage, thus it is much slower, requiring the observer to
scan each element of the image serially.

It is interesting to note that the FIT resembles a broader category of
paradigms, namely {\it the dual process models} \cite{Stanovich2000}.
Under this conceptual framework, complex cognitive tasks usually
consist of two systems, that essentially differ in which Kahneman
\cite{Kahneman2003} referred as (1) effortless intuition, and (2)
deliberate reasoning. The system (1) comprises processes that are
fast, intuitive and can be performed automatically and in parallel,
like when trying to identify the state of spirit of a person based on
her/his facial expression. These processes are acquired through habit,
being usually inflexible and hard do control or modify.  The system
(2), on the other hand, is characterized by slow, serial but extremely
controlled processes, which is the case, for example, when one tries
to solve a mathematical equation.

The FIT was thoroughly studied and expanded during subsequent years
\cite{Quinlan2003}. Although regarded, in its initial form, as an
oversimplification \cite{Treisman1990}, it surely represents a
formidable conceptual starting point point for research on the
subject. Two particular assumptions of the FIT, namely that the whole
optical array is homogeneously analyzed and that attention can be
displaced independently of the eye movements (thus being called {\it
  covert attention}), are of interest to be expanded upon
\cite{Underwood1998b}. It is widely known that visual acuity falls
rapidly from the point of fixation \cite{Anstis1974}, being confined
to a small region called fovea. While there is little doubt about the
existence of covert attention \cite{Hoffman1995}, it has been argued
that situations in which covert attention performs better than overt
eye movements are unusual and restricted to laboratory tests
\cite{Underwood1998b}. These led to further investigation about the
function of eye movements in visual search
\cite{Viviani1982,Findlay1995,Findlay1999}.

Eye movements are composed of fixations and saccades, but even during
fixations, the eyes are not completely still. In fact, fixational eye
movements (FEyeM) include drift, tremor and microsacades. The drift
corresponds to the erratic and low velocity component of FEyeM. The
	tremor is irregular and noise-like with very high frequency, while
microsacades correspond to small rapid shifts in eye position akin to
saccades, but preferentially taking place on horizontal and vertical
directions \cite{Rolfs2009}. Whether or not each one of these
movements play an effective role in visual cognition still represents
a rather controversial issue \cite{Engbert2004,Engbert2006,Engbert2011}, 
but it is known widespread that, if the FEyeM halt, visual perception
stops completely. Previous attempts to model eye movements have been
mainly devoted to describe the sequence of fixations and saccades in
terms of stochastic processes \cite{Crutchfield1998} like regular
random walks \cite{Matin1970}.  Very often, the gaze is considered as
a random walker subjected to a potential extracted from a saliency
map, namely a field that depends on the particular features of the
image under inspection, such as color, intensity and orientation
\cite{Itti1998,Brockmann1999,Parkhurst2002,Boccignone2004}.

Recent research on visual cognition has been directed to the
development of experimental and analytical methods that can
potentially elucidate the interplay between different components of
cognitive activities, and how their interactions give rise to
cognitive performance \cite{Holden2009}. While the detection and
integration processes mentioned before represent basic components of
visual cognition that can be investigated separately, the way they
interact should be relevant for the comprehension of more intricate
visual tasks. Therefore, it is of paramount interest to determine if
cognitive dynamics is dominated by components or interactions. Here we
show through eye-tracking experiments that the cognitive task of
visual search for hidden objects displays typical statistical
signatures of interaction-dominated processes. Interestingly, by
increasing the difficulty level of the visual task, our results also
indicate that the eye movement changes from a serial reading-like
(systematic) to an isotropic (random) searching strategy.

\begin{figure*}[h]
    \begin{center}
        \includegraphics[scale=0.14]{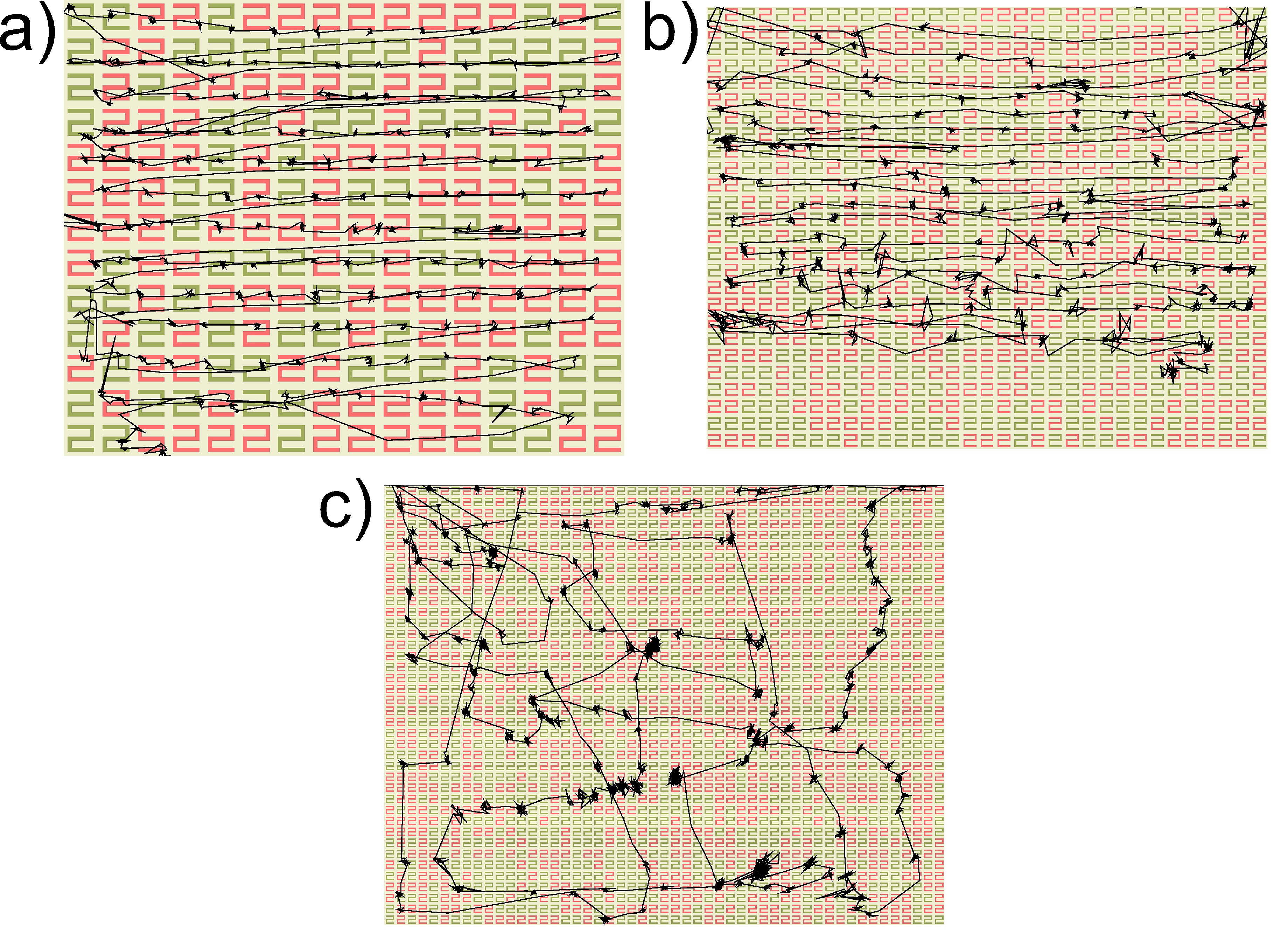}
    \end{center}
  \caption{{\bf Search over {\it 5-2} lattices.} The subjects try to
    find a single number {\it 5} in an array of red and green numbers
    {\it 2} (distractors). Of course, the larger the size of the array
    (number of distractors), the more difficult is the searching task.
    Two distinct types of searching patterns are clearly observed. The
    systematic search shows an anisotropic characteristic, as in (a)
    and (b), with the eye moving more frequently in a particular
    direction, horizontally most often, but also vertically for some
    subjects. In the case of random search, as in (c), the eyes are
    likely to move equally in any direction. Our results also show
    that the frequency with which the subjects follow the systematic
    pattern decreases with the difficulty of the test. Two thirds of
    the recordings (42 out of 63), in the case of difficulty 0,
    correspond to systematic searches, while only half of the
    recordings (16 out of 32) displayed this behaviour with tests of
    difficulty 1. In the presence of a large number of distractors,
    most of the subjects prefer to follow a random search strategy.
    This is the case with tests of difficulty 2, where only one fourth
    of the recordings (6 out of 24) show systematic behaviour.}
    \label{fig:52_gaze}
\end{figure*}

\section{Results}

\begin{figure*}[b]
\begin{center}
    \includegraphics[scale=0.25]
        {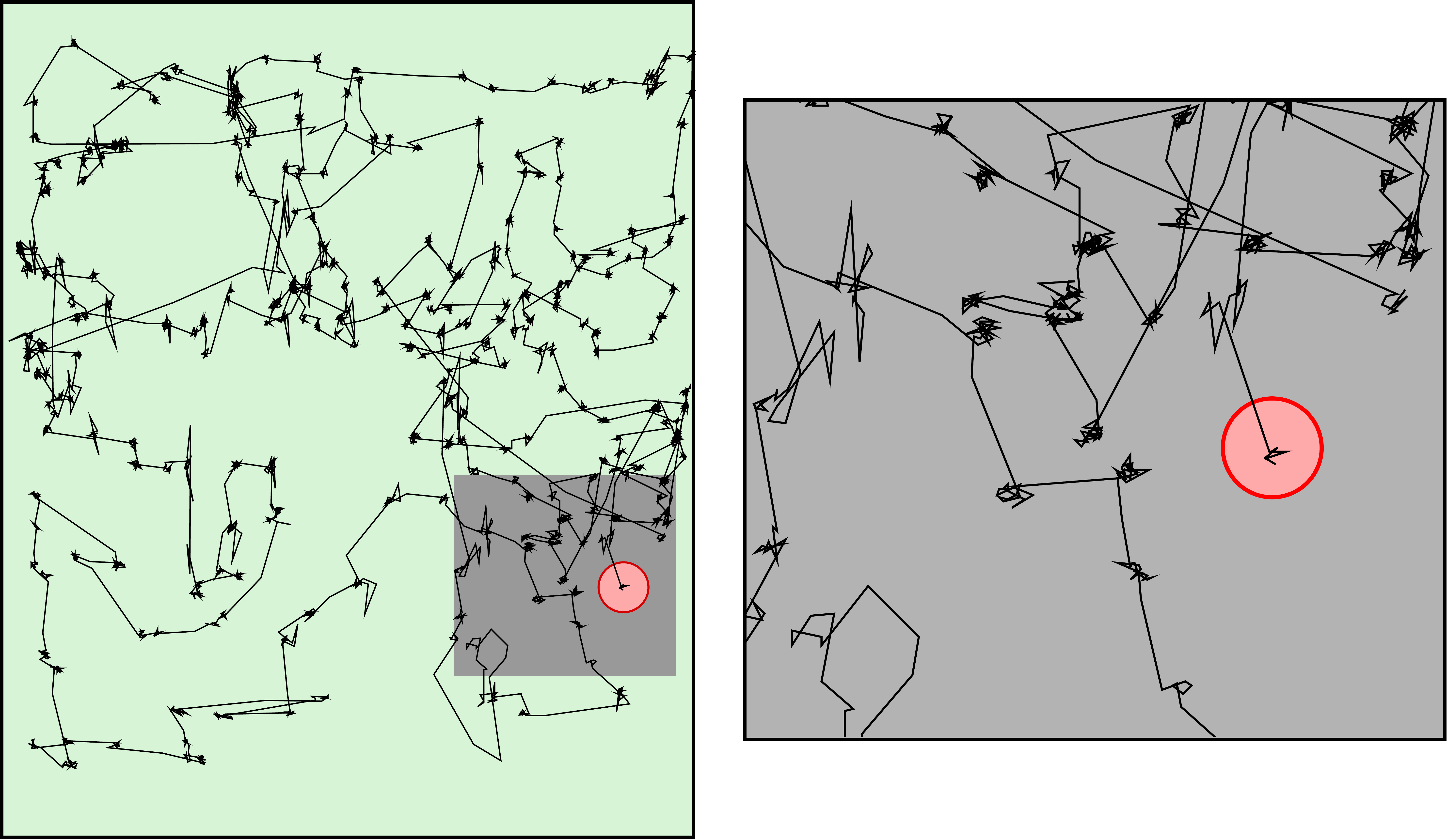}
    \end{center}
  \caption{{\bf Where's Wally?} On the left, we show the result of a
    typical eye-tracker recording of a search task on a complex
    landscape~\cite{Wally1}. The basic elements of eye movement are
    clearly present, namely the numerous sets of fixation points
    connected by large jumps (saccades)~\cite{Underwood1998a}. In this
    particular searching test, the points of fixation bunch up around
    some regions, where certain details of the image demand more
    attention than others, however, one can not perceive any
    systematic pattern in the trajectory of the eyes. On the right,
    part of the image is enlarged, where the red circle indicates the
    location of the target. The recording process ends when the target
    is found. \label{fig:wally_gaze}}
\end{figure*}

Visual search experiments have been performed with targets hidden in
two different types of disordered substrate images (see {\it Methods}
for details). In the first, as depicted in Fig.~\ref{fig:52_gaze}, the
subjects were asked to search for a target (number {\it 5}) in an
image with distractors (numbers {\it 2}) placed on a regular array.
Figure~\ref{fig:wally_gaze} shows an example of the second type of
test, where we utilized images from the book series ``Where's Wally?''
\cite{Wally1}. These last can be considered as very complex images,
since distractors are irregularly placed in an off-lattice
configuration and specially drawn to closely resemble the target. The
resulting image designed under these conditions frequently leads to a
searching task of enhanced difficulty.  The analysis of the results
from the two tests enabled us to identify general statistical patterns
as well as particular features in the eye movement that are related
with the irregularity and complexity of the image adopted in the
eye-tracking experiments.

In the case of the {\it 5-2} lattice tests, the typical trajectories
shown in Fig.~\ref{fig:52_gaze} indicate that, when the number of
distractors is small, most subjects performed systematic searches,
that is, the task is accomplished in a manner that resembles a person
reading a text, for example, from left to right and/or from top to
bottom. By increasing the number of distractors, a transition can be
observed from this directional (systematic) trajectory to an isotropic
random strategy of searching for the large majority of the
experiments.  Precisely, systematic patterns have been observed in two
thirds of the eye-tracking recordings (42 out of 63) for difficulty 0
(see Fig.~\ref{fig:52_gaze}a), half of the recordings (16 out of 32)
for difficulty 1 (see Fig.~\ref{fig:52_gaze}b), and only one fourth of
the recordings (6 out of 24) for difficulty 2 (see
Fig.~\ref{fig:52_gaze}c). No discernible systematic searches were
observed in the case of ``Where's Wally?'' tests.

\begin{figure*}[h]
\begin{center}
    \includegraphics[scale=0.35]
        {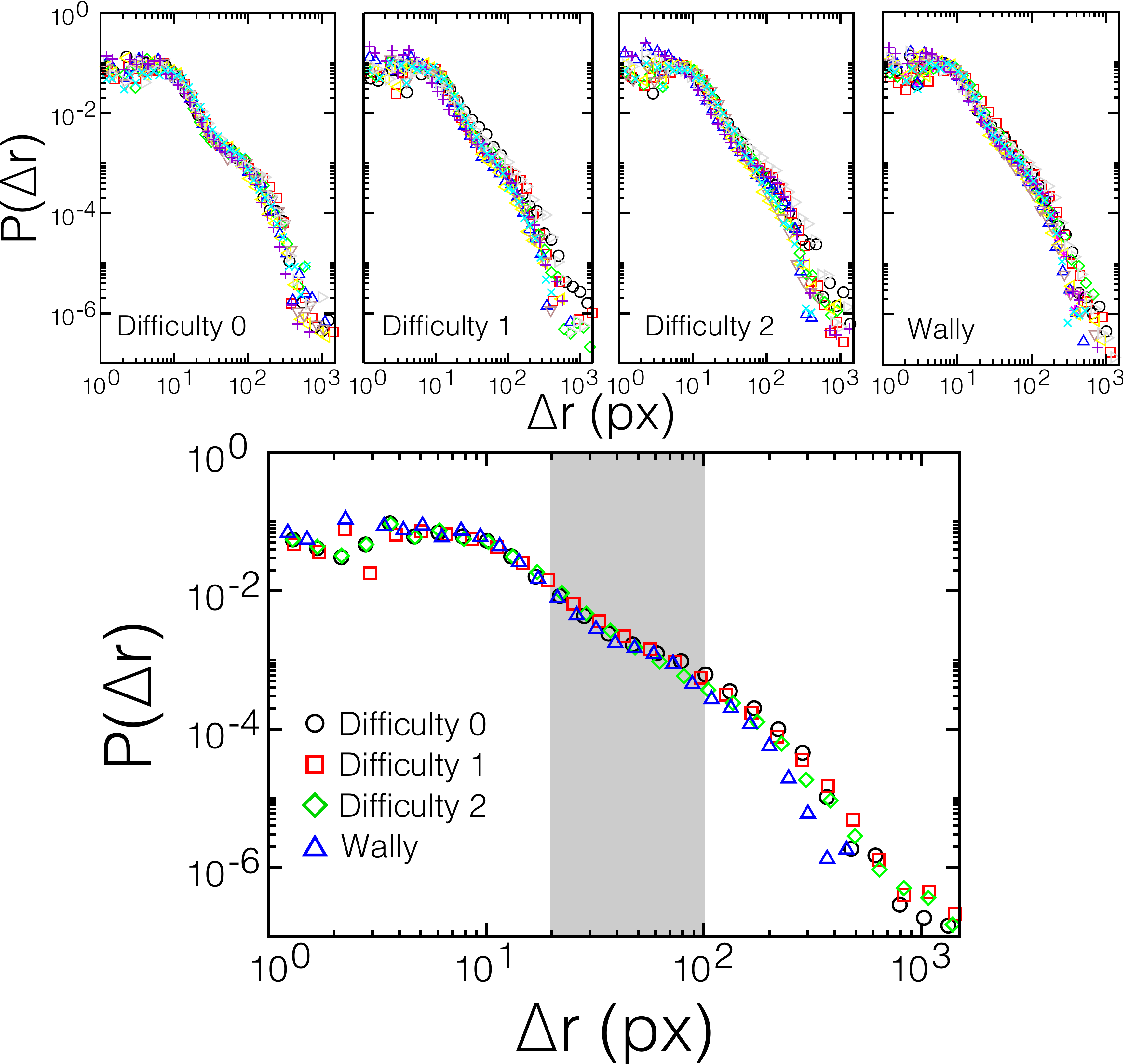}
    \end{center}
  \caption{{\bf Log-log plots of the distributions of gaze jump
      sizes.} The size of a jump corresponds to the distance, measured
    in number of pixels, covered by the eye gaze in a interval of
    17~milliseconds. The logarithmic plots on the top are the
    distributions of jump sizes obtained for different subjects and
    each of the tests performed, namely ``Where's Wally?'' and {\it
      5-2} lattices with difficulties 0, 1 and 2.  Despite small
    variations, all distributions show a similar quantitative
    behaviour. This ``universal'' statistical signature of the
    searching process can also be detected from the results displayed
    on the bottom panel, where the distributions of gaze jump sizes
    averaged over all subjects are shown.  The similar shape observed
    in all distributions suggests that identical mechanisms control
    the amplitude of the gaze shift, regardless of the systematic or
    random aspects of the searching movements on the {\it 5-2}
    lattices, and the distinctive features of the underlying
    arrangement of distractors that compose the ``Where's Wally?''
    landscapes. The shaded area delimits a depression region that
    appears systematically in all distributions, where the sizes of
    fixational and saccadic eye movements overlap.}
        \label{fig:all_jumps}
\end{figure*}

Next, we analyze the size distributions of gaze jumps calculated for
the raw data obtained from eye-tracking experiments. By definition,
the size of a jump in this case corresponds to the distance, measured
in number of pixels, covered by the eye gaze during each recording
step of the eye-tracker device, adjusted here for approximately
17~milliseconds. Strikingly, as depicted in Figs.~\ref{fig:all_jumps},
all tests produced alike distributions of gaze jumps, regardless of
the subjects, complexity of the tests, or the search strategy (regular
or random). This universal shape reflects the fixation-saccade duality
of the eye movement and clearly points to a superposition of
behaviours instead of a description in terms of pure monomodal
distributions \cite{Stephen2009,Bogartz2012}.

\begin{figure*}[h]
\begin{center}
    \includegraphics[scale=0.35]
        {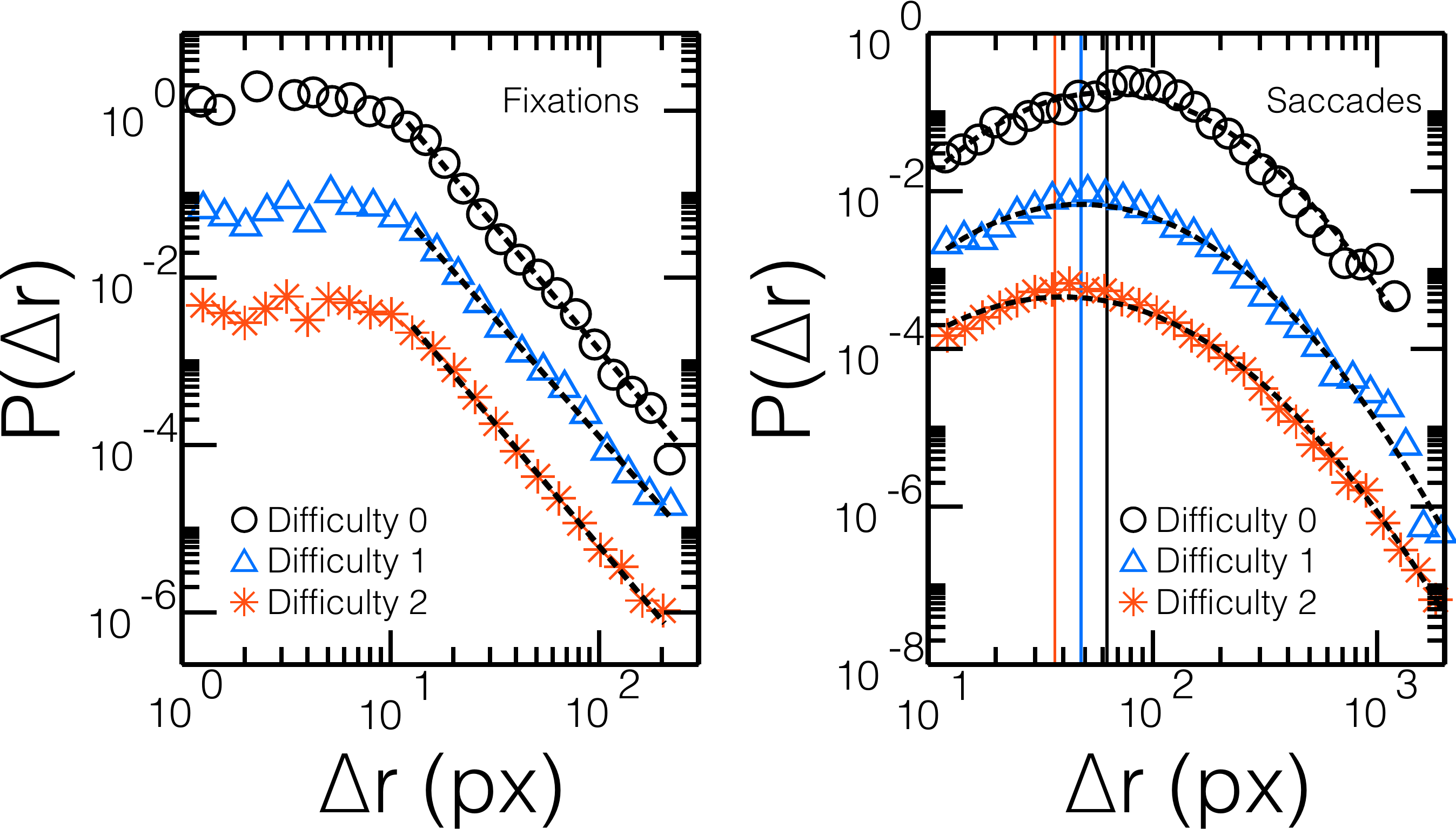}
    \end{center}
  \caption{{\bf Jump sizes distributions of filtered data for the {\it
        5-2} searching tasks.} For each difficulty level, the
    corresponding data has been averaged over all subjects.  By
    applying the filter developed by Olsson \cite{Olsson2007} to the
    raw data, we can distinguish between fixational and saccadic
    movements, so that the jump size distributions for each mechanism
    of eye movement can be studied separately. For better
    visualization, the resulting distribution curves are shifted
    vertically by a factor of $1/16$ and $16$, in the case of
    difficulty 0 (black) and difficulty 2 (orange), respectively. As
    depicted, the tails of the fixation curves can be adequately
    fitted by power-laws, $P(\Delta r) \propto \Delta r^{-\alpha}$
    (dashed lines), with an exponent, $\alpha\approx 2.9$, for all
    tests (see Table \ref{tab:mmse}). The distributions of jump sizes
    for the saccadic movements follow a quite different behaviour,
    which is compatible with a log-normal distribution, as the best
    fits to the three data sets show (dashed lines) on the left panel.
    The corresponding fitting parameters are presented in
    Table~\ref{tab:mmse}. Interestingly, the most frequent length of
    the saccades (arrows) decreases with the difficulty of the test.
    This could possibly happen either because the distractors (numbers
    {\it 2}) are simply smaller in the more difficult tests or the
    saccades are influenced by the colors of the distractors (that
    form relatively smaller clusters in more difficult settings), or a
    conjunction of both effects.
      \label{fig:52_jumps}}
\end{figure*}

\begin{figure*}[h]
\begin{center}
    \includegraphics[scale=0.5]
        {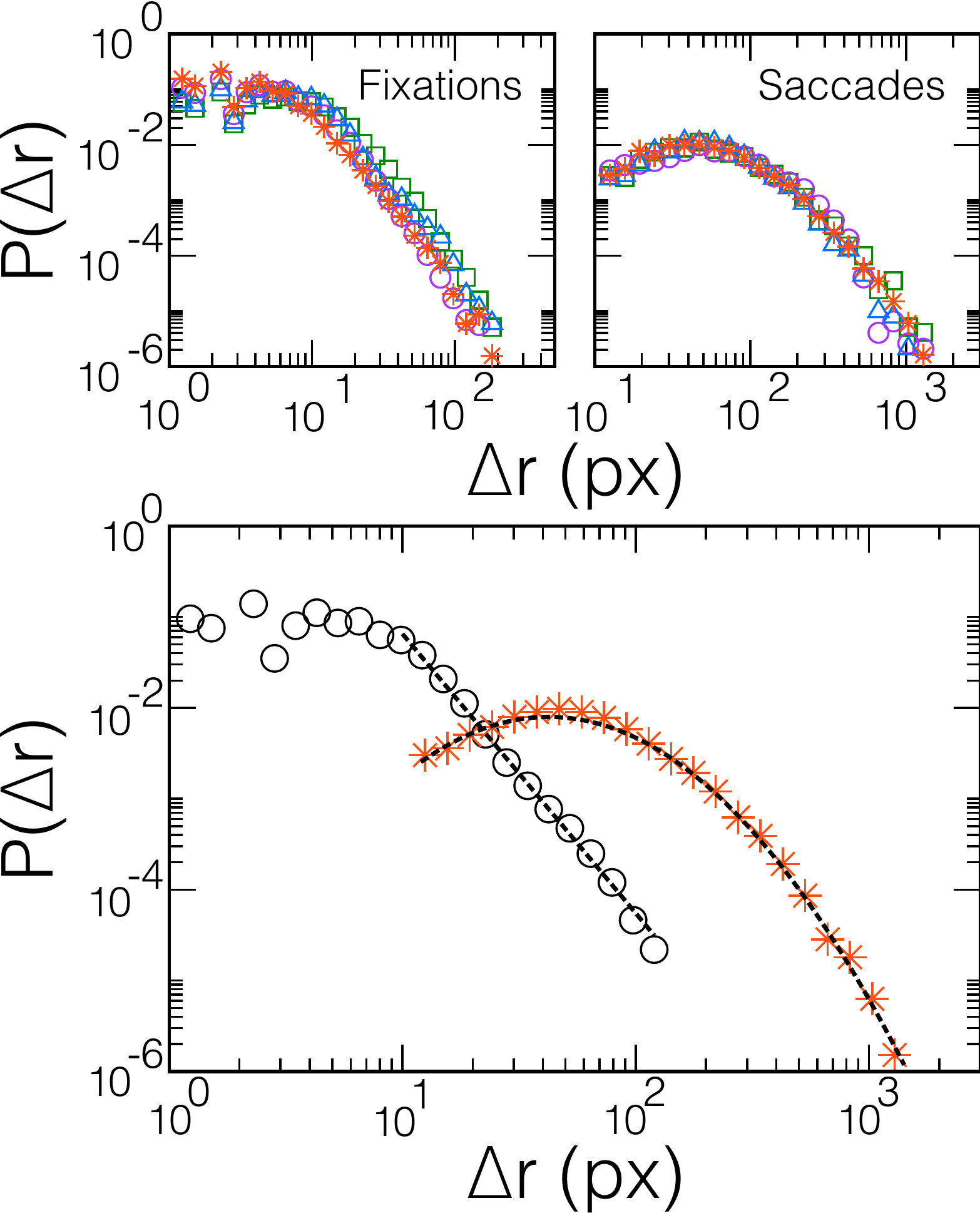}
    \end{center}
    \caption{{\bf Jump sizes distributions of filtered data for the
        ``Where's Wally?''  searching tasks.} The top panels on the
      left and right show the results for fixational and saccadic gaze
      jumps, respectively, calculated for different subjects. On the
      bottom, we show the same distributions, but now averaged over
      all subjects. The dashed lines correspond to the best fits to
      the data sets of power-laws, for the fixational movements, and
      log-normal distributions, for the saccadic movements. The
      statistical features of both mechanisms of eye movement observed
      here are quite similar to the ones identified for the {\it 5-2}
      searching tests (see Fig~\ref{fig:52_jumps}).
      \label{fig:wally_jumps}}
\end{figure*}

The presence of two modes separated by a slight depression that marks
the overlap region can be observed in practically all jump size
distributions of the raw data. Such a behaviour strongly suggests the
need for a filtering process through which fixations and saccades can
be adequately identified and their statistical properties
independently studied. With this purpose, here we apply a modified
version of the fixation filter developed by Olsson \cite{Olsson2007},
as described in the {\it Methods} section. As shown in
Figs.~\ref{fig:52_jumps} and \ref{fig:wally_jumps}, the resulting
distributions of jump sizes for fixational movements obtained for {\it
  5-2} and ``Where's Wally?'' tests, respectively, also display the
same statistical signature. Precisely, for gaze steps larger than
$10\,px$, the distances $\Delta r$ follow typical power-law
distributions,
\begin{equation}
	P(\Delta r) \propto \Delta r^{-\alpha},
	\label{eq:powerl}
\end{equation}
with a statistically identical exponent, $\alpha \approx 2.9$, for all
tests (see Table \ref{tab:mmse}). For gaze steps smaller than
$10\,px$, the distributions display approximately uniform behaviour,
possibly due to the fact that, in this scale, eye tremor is of the
order of drift, although this hypothesis cannot be tested with the
time resolution used in our measurements \cite{Engbert2003}.  Once
identified through the filtering process, the analysis of the saccadic
movements in all tests reveals that the distributions of sizes for
this type of eye jump can be well described in terms of a log-normal
distribution,
\begin{equation}
    P(\Delta r) = \frac{1}{\Delta r\sqrt{2\pi\sigma^2}}\exp\left[{-\frac{
    \left(\log \Delta r - \mu\right)^2}{2\sigma^2}}\right],
    \label{eq:lognormal}
\end{equation}
where the parameters $\mu$ and $\sigma$ correspond to the average and
variance of the logarithm of the saccade length, respectively. Once
more, the fact that a single distribution function can properly
describe the general statistical features of different searching tests
suggests that same underling mechanisms control the cognitive task
under investigation. It is interesting to note, however, that the
numerical values of the estimated parameters of the distributions
depend on the details of the test. For instance, in the case of {\it
  5-2} tests, the mode of the distribution (the most probable length)
decreases systematically with the difficulty of the searching task,
indicating that saccadic movements somehow adapt to the complexity of
the image.

\begin{table}[h]
\begin{centering}
\begin{tabular}{cc|cc}
 & \multicolumn{1}{c}{Fixations} & \multicolumn{2}{c}{Saccades}\tabularnewline
\cline{2-4} 
 & $\alpha$ & $\mu$ & $\sigma^{2}$\tabularnewline
\hline
\hline 
Difficulty 0 & $2.951\pm0.054$ & $4.810\pm0.050$ & $0.675\pm0.415$\tabularnewline
Difficulty 1 & $2.825\pm0.020$ & $4.594\pm0.018$ & $0.727\pm0.168$\tabularnewline
Difficulty 2 & $2.938\pm0.019$ & $4.510\pm0.019$ & $0.830\pm0.173$\tabularnewline
Wally & $3.091\pm0.017$ & $4.444\pm0.012$ & $0.698\pm0.112$\tabularnewline
\end{tabular}
\par\end{centering}
\caption{{\bf Parameters of the jump size distributions obtained from
    the non-linear least squares
    fitting to the filtered data.} The fixational steps follow
  a power-law, $P(\Delta r) \propto \Delta r^{-\alpha}$, for $\Delta
  r>10\, px$,
  while the saccadic jumps display a log-normal type of behaviour,
  $P(\Delta r)=\exp[-(\log \Delta r -\mu)^2/2\sigma^2]/\Delta
  r\sqrt{2\pi\sigma^2}$.
  The error represent a bootstrap estimation of the 95\% confidence
  interval \cite{Felsenstein1985}.
    \label{tab:mmse}}
\end{table}

\section{Discussion}

In summary, our results from eye-tracking tests in which subjects are
asked to find a specific target hidden among a set of distractors (see
{\it Methods}) reveal a gradual change on the searching strategy, from
a directional reading-like (systematic) to an isotropic (random)
movement as the number of distractors increases. However, regardless
of the differences in image complexity, searching tasks and individual
skills of the subjects, we observe universal statistical features
related with the distributions of gaze jump sizes. These distributions
generally show a characteristic bimodal behaviour, consequence of the
intrinsic dual nature of eye movement \cite{Bogartz2012}, that
alternates between saccades and fixations.

The application of a fixation filter to the raw data enables us to
study separately the distributions of jump sizes for fixational and
saccadic gaze steps. We find that the distribution of fixational
movements show long tails which obey power-laws \cite{Sims2007}, while
saccades, on the other hand, follow a log-normal type of behaviour.
The fact that both log-normal and power-law distributions arise from
multiplicative processes \cite{West1989} provide strong support to the
hypothesis that the interactions between components dominates the
cognition task of visual search \cite{Stephen2009}. In a dynamics
governed by interactions, the organization of the components and the
way they process information are context dependent, with no particular
function being encapsulated in any of the components themselves. This
non-linear response to the influx of information would give rise to
multiplicative distributions like the ones we disclosed here.

These observations are in evident contrast with a component based
scenario, where the final performance of a given cognitive task
results from the simple addition of sub-tasks that usually process
information in a specialized manner. Instead of log-normal or
power-law distributions, a process like this would give rise to
Gaussian or other additive distributions (e.g., exponential or gamma
distributions) \cite{Stephen2010}. It is worth noting that our results
are conceptually consistent with previous studies describing complex
behaviour in visual cognition \cite{Aks2002,Orden2003,Shinde2011,Bogartz2012}.
As a perspective for future work, it would be interesting to relate our
findings with other potential approaches based on non-cognitive random
strategies, where the searching task can be the result of an optimization
process \cite{Viswanathan1999,Snider2011,Benichou2011,Najemnik2005}.

\section{Methods}

\subsection{Equipment}

Eye movements were recorded with a Tobii T120 eye-tracking system
(Tobii Technology). In this study we only consider data obtained after
a valid calibration protocol is applied to both eyes of the subject.
The stimuli were presented on a 17'' TFT-LCD monitor with resolution
$1024\times1280$ pixels and capture rate of $60$~Hz.

\subsection{Tests}
Two types of tests consisting of visual searching for a hidden target
randomly placed among a set of distractors were performed by 11
healthy subjects with an average age of 23 years. The stimuli of the
first test consists of a square lattice composed of a single target
number {\it 5} and several number {\it 2}'s serving as distractors.
All numbers (target and distractors) are randomly colored red or
green, hindering the visual detection of the target through the
identification of patterns on the peripheral vision. This images were
organized in three difficulty levels according to the number of
distractors, labeled 0, 1 and 2 for 207, 587 and 1399 distractors,
respectively.

The stimuli of the second test are scanned images from the ``Where's
Wally?'' series of books \cite{Wally1}. The complexity of these
images, where a large number of distractors (background characters)
are irregularly placed together with Wally, the hidden target
character, explains the high difficulty involved in this visual
searching task.  Not all images used had an actual target, since we
had no intention to track the time taken to find the target. Instead,
our objective was to induce the subjects to perform the searching task
as naturally as possible.

In order to stimulate subjects to search efficiently, in all tests,
they were told to have a limited time to find the target, but not
informed exactly how much time would be available. In the case of the
{\it 5-2} lattice tests, $1$, $1.5$ and $2$ minutes were given to
search the target for the difficulties 0, 1 and 2, respectively. For
the ``Where's Wally'' tests, the subjects had 2~min. A summary of the
parameters can be found in Table \ref{tab:b52_param}.

\begin{table}[h]
    \begin{center}
        \begin{tabular}{>{\centering}m{1.0cm}>{\centering}m{1.5cm}
                        >{\centering}m{1.5cm}>{\centering}m{1.5cm}
                        m{1.5cm}}
            \hline 
            Difficulty & \# of Images & \# of Distractors &
            Size of Distractors & Time Available
            \\ \hline\hline 
            0 & 8 & $16\times13$ & $76\, px$ & $1\, min$
            \\ \hline 
            1 & 4 & $33\times26$ & $38\, px$ & $1\, min\,30\, s$
            \\ \hline 
            2 & 3 & $40\times35$ & $25\, px$ & $2\, min$
            \\ \hline
            Wally & 5 & -- & -- & $2\, min$
            \\ \hline
        \end{tabular}
        \caption{Parameters used in the {\it 5-2} and ``Where's Wally'' search tasks.
                 \label{tab:b52_param}}
        \par
    \end{center}
\end{table}

\subsection{Fixation Filter}
We adopted a modified version of the fixation filter developed by
Olsson \cite{Olsson2007} in order to identify which gaze points belong
to fixations and which belong to saccades. The basic idea is to
distinguish between segments of the signal that are moving slowly due
to drift, thus identified as part of a fixational sequence, from those
moving faster, constituting the saccades. This is achieved here by
taking the raw signal output, $\mathbf{s}_i$, namely the position of
the gaze captured at each timestamp $i$, and calculating for each
point the mean position of two sliding windows of size $r$, one
retarded and the other advanced,
\begin{equation}
    \label{eq:mbefore}
    \mathbf{m}^{\pm}_i = \frac{1}{r}\sum^{r}_{k=1}\mathbf{s}_{i\pm k}.
\end{equation}
The distance between them is calculated as,
\begin{equation}
    \label{eq:dist}
    d_i = \left|\mathbf{m}^{+}_i - \mathbf{m}^{-}_i\right|.
\end{equation}
Since each timestamp has the same duration, the displacement given by
Eq.~\ref{eq:dist} may be analyzed in the same way as the average
velocity, thus if $d_i$ is larger than its two neighbors ($d_{i-1}$
and $d_{i+1}$), and is also larger than a given velocity threshold, it
is considered a peak. If two peaks are found within the interval of a
single window, only the largest one is considered.

At this stage, the gaze points are divided into clusters separated by
the peaks. In the original filter \cite{Olsson2007}, the median
position of each cluster is used to locate the corresponding fixation.
Since we are instead interested in separating the gaze points that
correspond to fixations from those that belong to saccades, the radius
of gyration for each cluster $C$ is then calculated as,
\begin{equation}
    \label{eq:gyrad}
    R_g = \sqrt{\frac{1}{N}\sum_{i\in C} \left|\mathbf{s}_i - \mathbf{\bar{s}}\right|^2},
\end{equation}
where $\mathbf{\bar{s}}$ is the mean position of the gaze points that
belong to $C$. Steps that fall inside the circle area covered by the
radius of gyration, and are centered at $\mathbf{\bar{s}}$, are
considered to be fixational. The same applies to those steps that
leave this area but return to it without passing through another
fixation cluster. All other steps are considered saccadic jumps.

\section{Acknowledgments}
  We thank the Brazilian Agencies CNPq, CAPES, FUNCAP and FINEP, the
  FUNCAP/CNPq Pronex grant, and the National Institute of Science and
  Technology for Complex Systems in Brazil for financial support.

\section{Author contributions}
All authors contributed equally to the paper.

\section{Additional information}
{\bf Competing financial interests:} The authors declare no competing
financial interests.

\end{document}